\documentclass[twocolumn,showpacs,preprintnumbers,amsmath,amssymb]{revtex4}
\usepackage{graphicx}
\usepackage{dcolumn}
\usepackage{bm}

\begin{document}

\title{The importance of thermal disorder and electronic occupation for the
 T-dependence of the optical conductivity in FeSi and MnSi.}

\author{T. Jarlborg}

\affiliation{
DPMC, University of Geneva, 24 Quai Ernest-Ansermet, CH-1211 Geneva 4,
Switzerland
}


\begin{abstract}

The spectral weight (SW) for optical transitions in FeSi and MnSi
are calculated as function of temperature by means of LMTO-LDA
band calculations. The main effects, caused by structural disorder and electronic Fermi-Dirac
distribution, act oppositely on the T-dependence of the SW, while the
variation of the magnetic moment in MnSi
has only a minor effect. The calculations agree 
with the experimental
findings of an increasing SW in FeSi and a decreasing SW in MnSi as function of T.
The results can be understood from the change of the bandstructure with disorder.

\end{abstract}


\maketitle


Binary compounds of the B20 structure, like MnSi, FeSi, FeGe, exhibit many
unusual properties. FeSi is known for a crossover from being a narrow-gap semiconductor
at low T to an exchange enhanced paramagnetic metal at room temperature \cite{jac}.
Band calculations agree on the fact
that there is a small gap of the order 6 mRy at low-T \cite{matt,grech}. Different mechanisms,
such as strong correlation \cite{fu}, Kondo screening \cite{anis,aepp} and thermal
disorder \cite{jar97,jar99}, have been proposed for a description of the high-T
behavior. 
The isoelectronic FeGe is ferromagnetic (FM), but with a strong pressure (P) dependence
and a possible suppression of FM at high P
\cite{ped}. MnSi has one electron less per formula unit, and it 
is FM with helical spin order \cite{pflei}. Many questions about the magnetic state
in MnSi are still unresolved, as is reflected by many inconsistencies between spectral data, 
magnetic moment and band
calculations  \cite{corti,carb}. 

Recent studies of MnSi with
the technique of X-ray absorption suggest a correlated excite state, and the
interpretation in terms of bandstructure
is complicated by the involvement of a core hole \cite{carb}. 
However, through high-resolution optical measurements, it has been possible
to detect T-dependent details of the bandstructure of several B20 compounds
\cite{schl,degio,dama,marel,mena,guri}. The band filling of the small gap in FeSi as function
of raising T has been observed, but with an unexplained missing intensity of the total
spectral weight (SW, the integrated optical conductivity), 
at least for energies up to a few eV. This has been taken
as an indication of exotic, Kondo-like and band-unlike, interactions in FeSi \cite{schl}.
Subsequent measurements
concluded that the weight is essentially conserved for energies below $\sim$ 0.4 eV \cite{degio}.
Complementary optical measurements in other B20 compounds show that the non-conservation of the
spectral weight is quite general, and with contrasting behavior in FeSi and MnSi \cite{dama}.
From the view of LDA bands it is not evident that the SW should change with temperature.
Previous band calculations, where a disordered supercell is modeling the effect of thermal disorder,
explain the band filling, 
transport properties and other observed T-dependencies in FeSi \cite{jar97,jar99}. But the
question about non-conservation of the SW was not addressed.

In this letter we present band results and simulation of optical conductivity, $\sigma(E,T)$, in FeSi and MnSi
with emphasis
on effets of thermal disorder and electronic Fermi-Dirac (FD) occupation. This is done
through LMTO (Linear Muffin-Tin Orbital \cite{lmto,bdj})
band calculations in the local density approximation (LDA) \cite{lda}
 for ordered and disordered 64-atom supercells (2x2x2 extension
of the elementary cell) as has been described previously \cite{jar99}. 
Zero-point motion (ZPM) is not considered in this work, although small effects
of ZPM might be present \cite{ped}. Omission of ZPM implies that by comparing
SW from ordered and disordered structures, one may slightly exaggerate effects
of thermal disorder. Effects of electron-phonon coupling or spinfluctuations 
could be of importance in systems
with large density-of-states (DOS) at the Fermi energy (E$_F$). Such effects, as well as effects of
thermal expansion, are ignored in
this work.

The self-consistent convergence is relatively slow for the large supercells and the calculations
are restricted to two supercell configurations, both containing 64 atoms. One cell is perfectly ordered ("low-T") and 
the other is disordered ("high-T") with a mean displacement
($u$, which is assumed to be equal for both atoms)
of the atomic positions of 0.019$\cdot a_0$, where $a_0$ is the lattice constant.
This corresponds roughly to a temperature of 500K according to the measured data in ref. \cite{voc},
and is slightly higher to what was estimated in ref. \cite{jar99}. By comparison,
the ZPM from the same data corresponds to displacements of about 0.006$\cdot a_0$.
Instead of searching for a continous T-variation of the SW we present the trends and reasons for
non-conserved SW in the different B20 compounds by comparing the low-T and high-T results.
The final bands are calculated with equivalent density of k-points for the ordered
and disordered structures, for 125 k-points within 1/8 of the Brillouin Zone for the former
case and 405 points with half of the zone for the latter case. Other details of the calculations
are described in ref. \cite{jar99}.

The optical conductivity $\sigma(E)$ is related to the imaginary part of
the dielectric constant \cite{ash}. 
\begin{equation}
 \sigma(E) \sim  \sum_{i,j,k} \frac{f(\varepsilon_j^k) - f(\varepsilon_i^k)}{\varepsilon_j^k - \varepsilon_i^k}
\delta(\varepsilon_j^k - \varepsilon_i^k-E) \cdot M^2
\end{equation}
The evaluation of this expression 
contains a summation of band energies, $\varepsilon$ ($i$ occupied,
$j$ unoccupied) and k-points ($k$). For MnSi there is also a summation over the two spins.
The matrix element $M$ is calculated in the dipole approximation \cite{jn}
\begin{equation}
  M = \sum_q \int_0^{S_q} \Psi_{q,\ell,m}(\varepsilon_i^k,r) \vec{p} \Psi_{q,\ell^{'},m^{'}}(\varepsilon_j^k,r) d^3r
\end{equation}
where $\Psi$ are the wave function for the corresponding band and k-point, which includes s,p and d-states,
and $S_q$ is the Muffin-Tin radius for site $q$. The angular part of $M$ leads to the dipole
selection rule. Only transitions where $\ell - \ell^{'} = \pm 1$ and $m - m^{'} = \pm 1,0$ are allowed, and the radial
part can be simplified further \cite{jn}. The spectral weight, SW, is here defined as the energy integral
of $ \sigma(E)$.

This method was applied recently to FM FeGe, and the sequence and the relative intensities
of $\sigma(E)$ agree favorably with experiment \cite{guri}, however, with two
important differencies. First, the energy scale needs to be shifted by several tenths of an eV. Secondly,
the experimental spectrum is far less structured than the calculated one. The latter can be attributed
to life-time broadening, which is not included in the calculation. Similar deviations can be expected for 
MnSi, FeSi and CoSi \cite{mena}. Therefore, we do not aim at a detailed comparison with experimental data, but we
concentrate on the relative difference in two calculations of $\sigma(E,T)$, where the computational
procedures are indentical except for difference in temperature.

A few mechanisms can be imagined to cause non-conservation of the SW. T-dependent FM moments
remove or add spin states near E$_F$, which could change the allowed transitions across E$_F$.
This effect is small, since the amplitude of local FM moments is found to be rather stable.
The T-dependence of the FD occupation is well-controled in the band approach. If the DOS at E$_F$ is
large (as in the FM compounds MnSi and FeGe) it can reduce a considerable part of the SW within an energy range of
$k_BT$. 
The effect on the dipole matrix element from disorder is more subtle. A close look at the
$\ell$-character of the bands in ordered and disordered structures reveals a small
transfer of d-states towards other $\ell$-states, when the structural disorder is important, 
as is shown in the Table. (An increase of f-character is found in calculations with disorder within an elementary 
8-atom cell.) Thus, dipole transitions are stronger
for increasing T. This remark concerns these B20 compounds because of their exceptionally large
Fe-d (or Mn-d) character of the bands within a few eV around E$_F$.
Disorder will also change the bandstructure. Simply speaking, this can be resumed into the 
effect of band broadening. A normal degenerate band will be multiply split, when it
is projected on the bandstructure of the disordered supercell.
Interband transitions between the splitted bands are equivalent to
Drude-like intraband transitions within the original band (across E$_F$), and the band broadening becomes
larger for increasing disorder.

\begin{table}[b]
\caption{\label{table1}Difference in number of occupied electrons per transition metal atom in ordered
and disordered FeSi and MnSi.}
\vskip 5mm
\begin{center}
\begin{tabular}{|l|c|c|c|c| }
\hline
B20 compound & Mn/Fe-s & Mn/Fe-p & Mn/Fe-d & Total Mn/Fe  \\
\hline \hline
MnSi  &  +0.010 & +0.012  & -0.028 & -0.006 \\
FeSi  &  +0.010 & +0.004  & -0.009 & +0.005 \\
\hline
\end{tabular}
\end{center}
\end{table}

First we present results for paramagnetic (PM) and FM MnSi, ordered and disordred. The calculation
for ordered, FM MnSi gives a moment of about 0.63 $\mu_B$ per f.u., with a uniform distribution of the moment
on all (Mn) sites. The average moment for structurally disordered MnSi decreases slightly, to 0.58 $\mu_B$
per f.u., but with a very unequal distribution. The highest local Mn moment is about
three times larger than on the Mn-site with the lowest moment. Thus, structural disorder is
accompanied by a large spin disorder. 
The two spin-DOS functions near $E_F$ differ more and more 
when the average moment exceeds $\sim$0.6 $\mu_B$ \cite{lj}.
The DOS of PM MnSi is quite different from the two FM bandstructures, 
where the exchange splitting is reduced to zero from
about 0.5 eV in the FM cases. The effect
of magnetism on $\sigma$ for the FM, disordered case is masked by additional
effects coming from disorder, as will be discussed
later.
Despite these differencies in the three cases of MnSi there is a suprising similarity
of the $\sigma$'s shown in fig. 1, all shown for a FD-temperature of about 300K.  The general shape and the absolute
levels are basically the same. It seems that the large number of Mn-d states near E$_F$ allow for
many transitions and the differencies in band structure are not sufficient for being seen
in $\sigma$. 
It should be noted that the PM and FM calculations describe very different electronic structures,
much more different than what can be expected to be the case for MnSi, when T increases from 0 to 300K.
The conclusion is that T-variations of FM is not the cause for the increasing
SW in MnSi.

\begin{figure}
\vskip -5mm
\includegraphics[height=8.5cm,width=8.5cm,angle=-00]{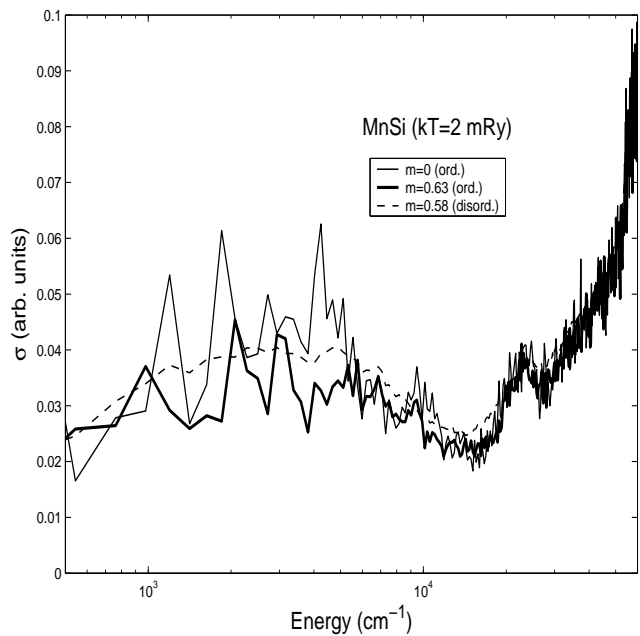}
\caption{Calculated variation of $\sigma$ for PM and FM MnSi
in ordered structure. The calculation for disordered structure
leads also to disorder of the amplitude of the FM local moment on Mn.
 }
\label{fig1}
\end{figure}

\begin{figure}
\vskip -5mm
\includegraphics[height=8.5cm,width=8.5cm,angle=-00]{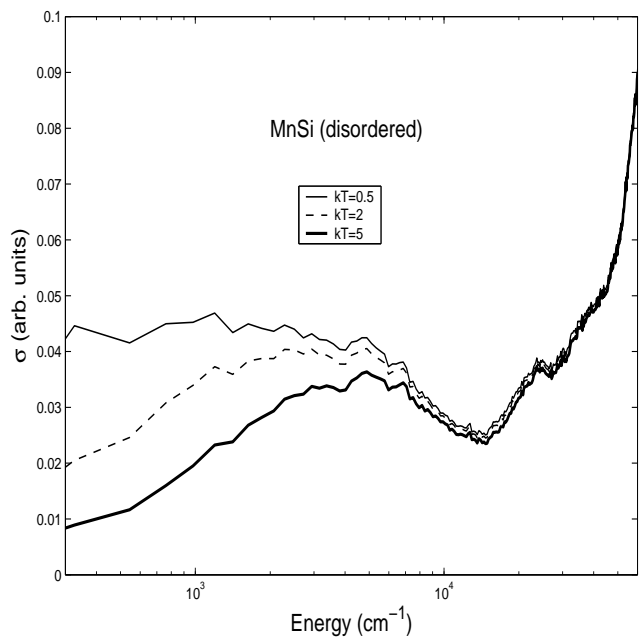}
\caption{Calculated variation of $\sigma$ for disorderd FM MnSi
for three different temperatures ($k_B T$ in mRy, 1 mRy = 158K) in the FD distribution.
The FD occupation reduce $\sigma$ at low energies for large T, and this
trend is general for all cases.
 }
\label{fig2}
\end{figure}

However, fig.2 shows that the T-dependent FD occupation has large effects
for $E = \hbar \omega$ below $\sim$ 1 eV. Such an effect is stronger in a material
with large $N(E_F)$, as in MnSi. The important transitions start at higher energies 
in FeSi and CoSi, because of their deep DOS minimum
near E$_F$, and the effect of FD occupation is smaller. 
The $E$-sum of $\sigma(E,T)$, the SW, evaluated at high T remains below the low-T curve,
also at large $E$, because of the loss of transitions at low $E$.
This mechanism explains the reduction of the SW for increasing T as is observed in MnSi \cite{dama}
and in FeGe above $\sim$ 0.1 eV \cite{guri}.
The same reason of transition probabilities explains why the SW-curve for a high-DOS material 
(MnSi, FeGe..) is above that for a
low-DOS material (FeSi, CoSi..).

The effects of disorder on $\sigma$ is two-fold in the case of FeSi. First, the bandstructure itself
will change because of the structural disorder, so that the small gap of about 6 mRy at low T will
be closed at intermediate T and completely filled at large T (see fig. 2 in \cite{jar99}).
This fact will not change the total SW. It will just "broaden" the bandstates, and redistribute
$\sigma$ to other energies $\hbar \omega$. 
Secondly, there are effects on the matrix elements. 
Fig. 3 shows $\sigma(\omega)$ for ordered and disordered
FeSi for equal T of the FD distribution. The first effect, the filling of the gap, is seen below
$\sim 1000 cm^{-1}$ (1/8th of an eV), but the curve for disordered FeSi ("high-T") seems to be above
the low-T curve even for quite large energies. The same observation can be made for MnSi in
fig. 1.
As was discussed above, the reason  
is that the dipole approximation allows for
increased transition probabilities at large disorder, because of the general trend of promoting
d-states into p- or f-states. The bandstructures are similar for all these 3d transition metal
compounds of B20 structure, and the calculated effects on the SW in PM FeSi and FM MnSi are about equal.

\begin{figure}
\vskip -5mm
\includegraphics[height=8.0cm,width=8.5cm,angle=-00]{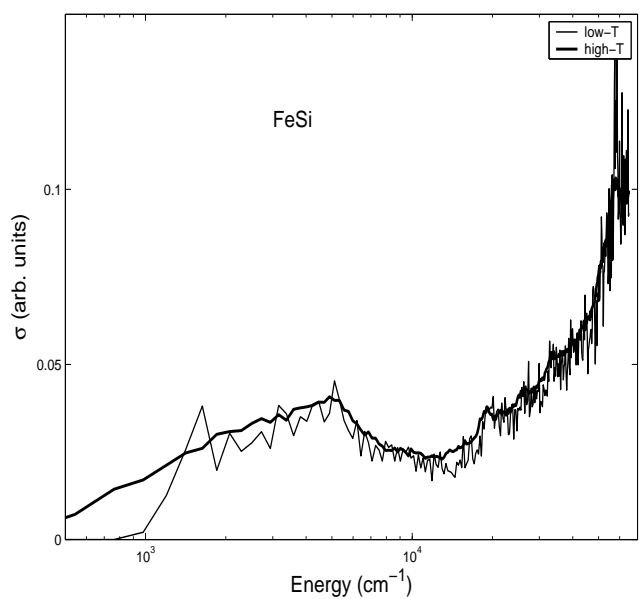}
\caption{Calculated variation of $\sigma$ for ordered ("low-T")
and disordered ("high-T") FeSi. The value of $k_B T$ in the FD
distribution is 1 mRy in both cases.
 }
\label{fig3}
\end{figure}

The effect of structural disorder on the matrix elements is the only mechanism which is found to increase
the SW for increasing T. The final result depends on the strength of the different mechanisms. 
The calculated SW at "low-T" (ordered structure and $T \sim 80$ K in the FD occupation)
and at "high-T" (disordered structure with $T \sim 500$ K in the FD occupation) 
for FM MnSi and (PM) FeSi are shown in fig. 4. The temperature of the lattice (structural disorder)
and of the electrons (FD occupation) are chosen to be equal. 
It is seen that the effect from the FD occupation is
dominating in MnSi, since the SW decreases at high T. The opposite behavior is seen for FeSi. First,
the low-T SW is zero below $\sim 0.1$ eV, and the high-T curve is non-zero because of the absense of the
gap. But the high-T curve is above the low-T curve for FeSi also at higher energies, mainly because of
matrix-element effects.  The opposite T-dependence in FeSi and MnSi agree with the
experimental findings \cite{dama}. It can be understood from the relatively low DOS
near E$_F$ in FeSi, which makes the effect from FD occupation weaker than the effect from disorder.
The lower DOS explains also why the SW-curves for FeSi is below those of MnSi.

The experimental SW in CoSi show almost no T-dependence \cite{dama}. This non-magnetic compound has one more electron
per transition metal compared to FeSi, and E$_F$ falls in a wide valley with very low DOS, but there
is no gap. The valley
is too wide to be filled by neighboring states for realistic disorder.  The effect of
increasing SW is smaller in CoSi than in FeSi, and it could lead to a cancellation of the two mechanisms, at least
for energies smaller than the width of the valley.


The recent measured SW in FeGe shows a decrease below $\sim$ 0.1 eV and an increase above this energy for
increasing T \cite{guri}. FeGe is isoelectronic with FeSi, but it is FM with a saturation moment of about 1 $\mu_B$
per f.u., and with a high DOS of the minority states. 
The majority band structure is similar to the one of CoSi, with E$_F$ in a wide DOS valley,
with only a moderate effect from the mechanism of FD occupation.
The T-dependencies at low $E$ should therefore mostly be caused by the minority bands, which are 
quite similar to those of PM MnSi. The high minority DOS at $E_F$ makes the process of FD occupation
efficient to cause a reduction of the SW at low $\hbar \omega$, while the effect from matrix elements 
should dominate for large $E$ and make the SW larger as function of T.

\begin{figure}
\vskip -5mm
\includegraphics[height=8.0cm,width=8.5cm,angle=-00]{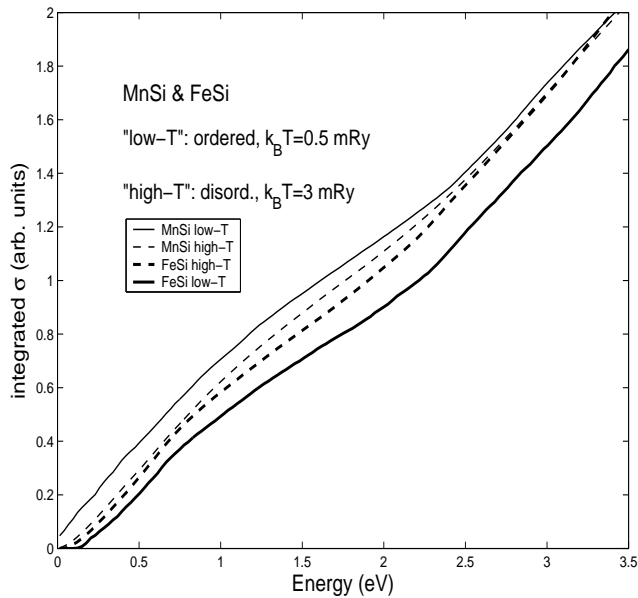}
\caption{Spectral weight of FM MnSi and PM FeSi at two temperatures,
see the text for the definitions of electron- and lattice-temperaure
. }
\label{fig4}
\end{figure}

In conclusion, it is shown that thermal effects from structural disorder 
and electronic excitations are important for an understanding of
the observed trends for the T-dependence of the spectral weight
in FeSi and MnSi. The two mechanisms work in general oppositely on the 
T-dependence.
Details of the bandstructures of FeSi and MnSi,
mainly the presence of a small gap in the former and a high DOS for the latter,
makes one or the other effect dominant. It is somewhat surprising to
find that PM or FM bandstructures in MnSi lead only to minor differences
in $\sigma$, even when disorder of the moments are taken into account.
These results lend further support to the idea that thermal disorder is the main cause
for the unusual T-dependent properties of FeSi.
The unique band
feature of a very narrow gap at E$_F$ makes this compound special in many respects.

 I am grateful to F. Carbone, V. Guritanu and D. van der Marel
for various discussions about the properties of B20 compounds.

\end{document}